\documentclass[letters,fleqn,usenatbib]{mnras}

\usepackage{newtxtext,newtxmath}
\usepackage[T1]{fontenc}

\DeclareRobustCommand{\VAN}[3]{#2}
\let\VANthebibliography\thebibliography
\def\thebibliography{\DeclareRobustCommand{\VAN}[3]{##3}\VANthebibliography}

\usepackage{graphicx}
\usepackage{amsmath}
\usepackage{hyperref}
\usepackage{orcidlink}
\usepackage{xspace}
\usepackage{xcolor}

\makeatletter
\AtBeginDocument{%
  \let\NAT@hyper@ORIG\NAT@hyper@
  \def\NAT@hyper@#1{%
    \in@{\hyper@natlinkbreak}{#1}%
    \ifin@
      \begingroup
        \def\hyper@natlinkstart##1{\Hy@backout{##1}}%
        \def\hyper@natlinkbreak##1##2{##1\hyper@linkstart{cite}{cite.##2}}%
        \NAT@hyper@ORIG{#1}%
      \endgroup
    \else
      \NAT@hyper@ORIG{#1}%
    \fi
  }%
}
\makeatother

\newcommand{\citekdm}{\protect\mbox{\protect\hyperlink{cite.katz2011}{KDM2011}}\xspace}
\newcommand{\citepaperI}{\protect\mbox{\protect\hyperlink{cite.klein2024a}{Paper I}}\xspace}
\newcommand{\citepaperII}{\protect\mbox{\protect\hyperlink{cite.klein2024b}{Paper II}}\xspace}
\newcommand{\citepaperIII}{\protect\mbox{\protect\hyperlink{cite.klein2024c}{Paper III}}\xspace}
\newcommand{\Jh}{\hat{J}}

\title[Librating EKL is a simple pendulum]{Hierarchical Three-Body Problem at High Eccentricities = Simple Pendulum\\IV: Octupole for Librating Kozai-Lidov Cycles}

\author[Y. Y. Klein]{
Ygal Y. Klein\orcidlink{0009-0004-1914-5821}\thanks{E-mail: ygalklein@gmail.com}
\\
Institute for Advanced Study, Einstein Drive, Princeton, NJ 08540, USA
}

\date{Accepted XXX. Received YYY; in original form ZZZ}

\pubyear{\the\year{}}

\begin{document}
\label{firstpage}
\pagerange{\pageref{firstpage}--\pageref{lastpage}}
\maketitle

\begin{abstract}
We solve analytically the long-term octupole evolution of \textit{librating} Kozai-Lidov cycles - those with a negative Kozai constant, in which the argument of pericenter librates - in the double-averaged restricted hierarchical three-body problem. Librating cycles reach extreme eccentricities when the normal component of the orbital angular momentum vanishes, just as rotating cycles do, but their slow dynamics was left unsolved: as noted by Katz, Dong \& Malhotra (2011), the azimuth of the eccentricity vector jumps by half a turn every cycle, so the leading-order octupole kick alternates in sign and cancels pairwise. We show that the surviving dynamics, at second order in the octupole strength, is a simple pendulum with explicit coefficients. The pendulum predicts slow oscillations of the normal angular momentum with amplitude linear in the octupole strength, evolving on a timescale of order the secular timescale divided by the octupole strength, together with an explicit criterion for orbital flips. At octupole strengths typical of hierarchical triples, these amplitudes are comparable to - and in part of the librating window can exceed - those of rotating cycles, provided one waits the correspondingly longer timescale. The analytic model agrees with numerical integrations of the double-averaged equations across the librating window, for octupole strengths typical of the hierarchical triples in which the eccentric Kozai-Lidov effect is studied - from hot-Jupiter formation to gravitational-wave sources.
\end{abstract}



\section{Introduction}

The secular evolution of a test particle that orbits a central mass and is perturbed by a distant companion - the restricted hierarchical three-body problem - consists of large, coupled oscillations of the eccentricity and the inclination, on timescales long compared with both orbital periods. When the equations of motion are averaged over both orbital periods (double averaged, DA) and the perturbing potential is truncated at its leading (quadrupole) multipole, the oscillations are periodic and admit an exact analytic solution: the Kozai-Lidov Cycles (KLCs) \citep{kozai1962,lidov1962}\footnote{For the early history of the problem, including the work of \citet{vonzeipel1910}, see \citet{ito2019}.}. Recently, the KLC itself was shown to be equivalent to a simple pendulum \citep{basha2025}. The next (octupole) order allows extremely high eccentricities and orbital \textit{flips} from prograde to retrograde motion for perturbers on an eccentric orbit (Eccentric Kozai Lidov, EKL) (\cite{katz2011}, hereafter referred to as \citekdm, \cite{ford2000,naoz2011,lithwick2011,naoz2013,sidorenko2018,lei2022}; for a review see \cite{naoz2016}). The EKL effect has been invoked in systems ranging from satellites and planets \citep{cuk2004,naoz2012,teyssandier2013,petrovich2015,anderson2016,vick2019,stephan2021,angelo2022,stephan2024wdkick,weldon2025,holzknecht2026} to compact objects \citep{antonini2012,antonini2014,hoang2018,stephan2016,liu2018,martinez2020,naoz2022emri,shariat2023wd,melchor2024,shariat2025merged,shariat2025lmxb,shariat2025cv,melchor2025,xuan2025}.

KLCs come in two classes, distinguished by the behaviour of the argument of pericenter $\omega$: it circulates for \textit{rotating} cycles and librates around $\pm\frac{\pi}{2}$ for \textit{librating} ones. For rotating cycles at high eccentricities, \citekdm derived averaged equations for the slow variables - the component $j_z$ of the dimensionless orbital angular momentum along the axis of the outer orbit, the azimuth $\Omega_e$ of the eccentricity vector around it and the Kozai constant that labels the two classes (all defined precisely in Section \ref{sec:coordinates}) - and in \citet{klein2024a} (hereafter \citepaperI) we showed that they reduce, for most initial conditions, to a simple pendulum with an explicit flip criterion; \citet{klein2024b} (hereafter \citepaperII) extended the model to include Brown's Hamiltonian \citep{brown1936a,brown1936b,brown1936c,soderhjelm1975,cuk2004,breiter2015,luo2016,will2021,tremaine2023,grishin2024a,lei2025a,lei2025b,gao2025} and \citet{klein2024c} (hereafter \citepaperIII) proposed that the resonance with a precessing quadrupole potential \citep{hamers2017,petrovich2017,klein2023,klein2024d} is also a simple pendulum - for rotating and librating cycles alike.

For the octupole, by contrast, the librating cycles were treated differently, for a clear reason. The octupole transfers angular momentum at the rate $\dot{j}_z\propto\sin\Omega_e$. On the rotating side $\Omega_e$ accumulates slowly, the transfer holds its sign over many consecutive cycles, and $j_z$ evolves at first order in the octupole strength - the drive behind the averaged equations and the flips. \citekdm showed that librating cycles lack this coherence: for them the azimuth changes by $\pi$ each cycle, the transfer reverses sign, and its accumulation cancels to leading order. They consequently focused on the rotating class, reporting for librating cycles significant modulation - including flips - only on much longer timescales (estimated there as $t\sim\epsilon^{-2}_\text{oct}t_\text{sec}$, where $\epsilon_\text{oct}$ is the dimensionless octupole strength and $t_\text{sec}$ the secular timescale, both defined in Section \ref{sec:coordinates}), which were not solved. The unsolved class is not a marginal one. At $j_z=0$ librating cycles reach the maximal eccentricity $e_\text{max}=1$ just like rotating ones, and the class is populated: ensembles that draw the initial argument of pericenter $\omega_0$ uniformly (a subscript or superscript $0$ denotes an initial value throughout) together with eccentric inner orbits - the rule in compact-object population syntheses - start an $O\left(1\right)$ fraction of their high-inclination systems on librating cycles (Section \ref{sec:discussion}). What the class lacked is not members but theory: the analytic treatments are built around the $\omega=0$ crossing that only rotating cycles possess (\citekdm, \citepaperI, \citepaperII, \citealt{sidorenko2018}, \citealt{lei2022}), and the numerical examples and parameter surveys guiding them are initialized at $\omega_0=0$ - selecting the rotating class by construction \citep[e.g.][]{lithwick2011,teyssandier2013}.

In this Letter we solve the librating side. The starting point is the geometric mechanism identified by \citekdm: every librating cycle the eccentricity vector crosses the vertical, its azimuth jumps by exactly $\pi$, and the octupole kick $\propto\sin\Omega_e$ alternates in sign from cycle to cycle and cancels pairwise - no evolution of $j_z$ survives at first order in $\epsilon_\text{oct}$, and the rate-format averaged equations of the rotating side do not exist here. We promote this observation to a quantitative per-cycle map and derive what survives the cancellation: at second order, the slow dynamics is again a simple pendulum - in the \textit{doubled} angle $2\Omega_e$ - whose coefficients we give explicitly. The pendulum predicts slow $j_z$ oscillations with amplitude $\propto\epsilon_\text{oct}$ and period $\sim1/\epsilon_\text{oct}$ KLCs - parametrically tighter and slower than their rotating-side counterparts, $\propto\sqrt{\epsilon_\text{oct}}$ and $\sim1/\sqrt{\epsilon_\text{oct}}$ (\citepaperI); and an explicit flip criterion - a window of initial azimuths whose width is set by the pendulum separatrix. We validate all predictions against numerical integrations of the DA equations.

\section{Coordinate system and slow variables}\label{sec:coordinates}

Consider a test particle orbiting a central mass $M$ on an \textit{inner} orbit with semimajor axis $a$ and eccentricity $e$ and a distant mass $m_\text{per}$ on an \textit{outer} orbit with $a_\text{per},e_\text{per}$ where $a/a_\text{per}\ll1$. Following \citekdm the \textit{z} axis is aligned with the (constant) outer orbital angular momentum and the \textit{x} axis with the pericenter of the outer orbit. The dynamics of the test particle are parameterized by the dimensionless angular momentum vector $\mathbf{j}=\mathbf{J}/\sqrt{GMa}$, where $\mathbf{J}$ is the orbital angular momentum per unit mass, and the eccentricity vector $\mathbf{e}$ (pointing to the pericenter); the two are orthogonal and their magnitudes $j$ and $e$ satisfy $j^2+e^2=1$. The polar angle $i_e$ and azimuth $\Omega_e$ of the eccentricity vector are defined by
\begin{equation}
    \mathbf{e}=e\left(\sin i_e \cos \Omega_e,\sin i_e \sin\Omega_e,\cos i_e\right).
\end{equation}
Expanding the perturbing potential to third (octupole) order in $a/a_\text{per}$ and averaging over both the inner and outer orbits (double averaging, DA) gives (e.g. \citekdm, \citealt{liu2015}, \citealt{petrovich2015}, \citealt{tremaine2023}, \citealt{luo2016})
\begin{equation}
\Phi_{\text{per}}=\Phi_{0}\left(\phi_{\text{quad}}+\epsilon_{\text{oct}}\phi_{\text{oct}}\right)\label{eq:phi_Per}
\end{equation}
where
\begin{equation}
  \phi_{\text{quad}}=\frac{3}{4}\left(\frac{1}{2}j_{z}^{2}+e^{2}-\frac{5}{2}e_{z}^{2}-\frac{1}{6}\right),\label{eq:phi_quad}
\end{equation}
\begin{equation}
  \phi_{\text{oct}}=\frac{75}{64}\left(e_{x}\left(\frac{1}{5}-\frac{8}{5}e^{2}+7e_{z}^{2}-j_{z}^{2}\right)-2e_{z}j_{x}j_{z}\right),\label{eq:phi_oct}
\end{equation}
\begin{equation}
  \Phi_{0}=\frac{Gm_{\text{per}}a^{2}}{a_{\text{per}}^{3}\left(1-e_{\text{per}}^{2}\right)^{\frac{3}{2}}},\,\,\,\,\,\,\,\epsilon_{\text{oct}}=\frac{a}{a_{\text{per}}}\frac{e_{\text{per}}}{1-e_{\text{per}}^{2}}. \label{eq:epsilon_oct}
\end{equation}
The evolution of $\mathbf{j}$ and $\mathbf{e}$ is governed by the \citet{milankovitch1939} equations
\begin{align}
\frac{d\mathbf{j}}{d\tau}&=\mathbf{j}\times\nabla_{\mathbf{j}}\phi+\mathbf{e}\times\nabla_{\mathbf{e}}\phi,\label{eq:milankovitch_j}\\
\frac{d\mathbf{e}}{d\tau}&=\mathbf{j}\times\nabla_{\mathbf{e}}\phi+\mathbf{e}\times\nabla_{\mathbf{j}}\phi,\label{eq:milankovitch_e}
\end{align}
where $\phi\equiv\Phi_{\text{per}}/\Phi_0=\phi_{\text{quad}}+\epsilon_{\text{oct}}\phi_{\text{oct}}$ is the dimensionless potential and time is expressed in units of the secular timescale $t_{\text{sec}}=\sqrt{GMa}/\Phi_0$ using $\tau\equiv t/t_{\text{sec}}$; these are the DA equations that all the numerical integrations of this Letter solve.

When $\epsilon_\text{oct}=0$ the potential is axisymmetric and the KLCs are classified by two constants: $j_z$, conserved by the axisymmetry, and the Kozai constant
\begin{equation}
C_{K}=\frac{4}{3}\phi_{\text{quad}} + \frac{1}{6} - \frac{1}{2}j^2_z=e^{2}-\frac{5}{2}e^2_z=e^{2}\left(1-\frac{5}{2}\sin^{2}i\sin^{2}\omega\right), \label{eq:CK}
\end{equation}
with $-\frac{3}{2}\leq C_K\leq1$, where $i$ is the orbital inclination (the polar angle of $\mathbf{j}$) and $\omega$ the argument of pericenter. The last form separates the two classes: wherever $\sin\omega=0$ it gives $C_K=e^{2}\geq0$, so cycles in which $\omega$ circulates - crossing $\omega=0$ every cycle - have $C_K>0$ (the \textit{rotating} cycles), while cycles with $C_K<0$ never reach $\sin\omega=0$, their $\omega$ librating around $\pm\frac{\pi}{2}$ (the \textit{librating} cycles). As in \citekdm and \citepaperI-\citepaperIII we focus on the high eccentricity regime $\left|j_z\right|\ll1$, and in this Letter on the librating class $C_K<0$. When $\epsilon_\text{oct}>0$ both constants become slow variables, evolving on timescales long compared to the KLC. The azimuth $\Omega_e$ is subtler: on the rotating side it is slow as well, precessing at a rate $\propto j_z$; on the librating side it jumps by $\pi$ every cycle already at $\epsilon_\text{oct}=0$, and the slow quantity is $\Omega_e$ modulo $\pi$ - equivalently the doubled angle $2\Omega_e$. At $j_z=0$ every KLC reaches $e_\text{max}=1$ exactly - librating and rotating alike (\citekdm): the two classes share the extreme-eccentricity regime; what separates them is precisely this behaviour of the azimuth, to which we turn next.

\section{The librating KLC: an exact pendulum with an azimuthal half-turn}\label{sec:monodromy}

This section treats the quadrupole limit $\epsilon_\text{oct}=0$, in which the cycles are exact KLCs, at $j_z=0$. The vector $\mathbf{j}$ oscillates along a fixed line in the \textit{x-y} plane and we may align the \textit{y} axis with it, so that $j_x=j_z=e_y=0$ at all times and $C_K=e^2_x-\frac{3}{2}e^2_z$ (same frame as \citepaperIII). In the librating class $C_K<0$ pins $e_z$ away from zero, while $e_x$ oscillates through zero. As shown in appendix A of \citepaperIII, the in-plane dynamics $\left(e_x,j_y,e_z\right)$ are an \textit{exact} simple pendulum (see also \citealt{basha2025} for the general statement at all $j_z$). With
\begin{equation}
    C_x = \frac{9}{8}\left(2C_{K}+3\right),\quad C_z = \frac{27}{8}\left(1-C_{K}\right),\quad k=\frac{C_x}{C_z}\in\left(0,1\right),
    \label{eq:CxCzk}
\end{equation}
the eccentricity cycle (one $e\to1$ passage to the next) is solved by Jacobi elliptic functions of parameter $k$, with period
\begin{equation}
    T_e = \frac{2K\left(k\right)}{\sqrt{C_z}},
    \label{eq:Te}
\end{equation}
where $K\left(m\right)$ is the complete elliptic integral of the first kind with parameter $m$ (the period of the full $\mathbf{j}$-line oscillation is $2T_e$, cf. \citepaperIII).

The property that controls everything that follows is a \textit{half-turn}, identified already by \citekdm: $\mathbf{e}$ is confined to the plane perpendicular to the $\mathbf{j}$-line, so $\Omega_e$ is constant during the cycle unless $\mathbf{e}$ crosses $\hat{\mathbf{z}}$ - which requires $e=e_z$ and is therefore possible only for $C_K<0$. For librating cycles this crossing happens once per eccentricity cycle: the in-plane components of $\mathbf{e}$ emerge reversed and the azimuth advances by exactly
\begin{equation}
    \Omega_{e,n+1}=\Omega_{e,n}+\pi
    \label{eq:monodromy}
\end{equation}
per cycle (at $j_z=0$, $\epsilon_\text{oct}=0$), where $\Omega_{e,n}$ denotes the azimuth at the $n$-th $e_\text{max}$ passage. Equation \ref{eq:monodromy} is exact and geometric - it does not depend on $C_K$ - and it has no rotating counterpart: on a rotating cycle $\mathbf{e}$ never crosses $\hat{\mathbf{z}}$, so the azimuth at $j_z=0$ is simply constant, and a finite $j_z$ adds only a slow continuous precession, at the quadrupole rate $\left\langle f_\Omega\right\rangle j_z$. Here $f_\Omega$ is the precession kernel of \citekdm - an explicit function of $i_e$, given there - and $\left\langle\cdot\right\rangle$ denotes, here and throughout, the time average over one cycle. On the librating side the azimuth is constant only modulo $\pi$ - it snaps, by half a turn, once per cycle - and a finite $j_z$ superposes the slow precession here as well.

\section{The alternating octupole kick}\label{sec:kick}

Up to leading order in $j_z$, the octupole changes $j_z$ at the rate $\dot{j}_z=-\epsilon_\text{oct}f_j\sin\Omega_e$, with the transfer kernel (\citekdm)
\begin{equation}
    f_j=\frac{75}{64}\,e\sin i_e\left[\frac{1}{5}-e^{2}\left(\frac{8}{5}-7\cos^{2}i_e\right)\right],
    \label{eq:fj_kernel}
\end{equation}
identical on the two sides of $C_K=0$. Integrating over one librating KLC at $j_z=0$ (during which $\sin\Omega_e$ is constant up to $O\left(j_z,\epsilon_\text{oct}\right)$) gives the per-cycle kick
\begin{equation}
    \Delta j_z = -\epsilon_\text{oct}\,\chi\left(C_K\right)\sin\Omega_{e},
    \label{eq:kick}
\end{equation}
where $\chi=T_e\left\langle f_j\right\rangle$ and the cycle average closes in elliptic form,
\begin{equation}
    \left\langle f_{j}\right\rangle =\frac{15}{64\sqrt5}\frac{\sqrt{3+2C_K}}{K\left(k\right)}\left[\left(4-11C_{K}\right)A\left(k\right) + 3\left(1-C_{K}\right)\sqrt{1-k}\right],
    \label{eq:fj_librating}
\end{equation}
with $A\left(k\right)=\arcsin\left(\sqrt{k}\right)/\sqrt{k}$. Equation \ref{eq:fj_librating} is the librating counterpart of the rotating average $\left\langle f_{j}\right\rangle$ of \citekdm (Equation 13 therein): the two expressions share the structural factor $\left(4-11C_{K}\right)$ and join continuously at $C_K\to0$. On the rotating side this factor produces the zero of the drive at $C_K=\frac{4}{11}$; on the librating side both bracket terms are positive and \textit{the drive never vanishes}.

The fork between the two sides is therefore not in $\left\langle f_j\right\rangle$ - it is in the azimuth equation that must accompany it. On the rotating side $\mathbf{e}$ never approaches $\hat{\mathbf{z}}$, so $f_\Omega$ is bounded along the cycle and averages to the slow precession rate $\dot{\Omega}_e=\left\langle f_\Omega\right\rangle j_z$; pairing it with $\dot{j}_z=-\epsilon_\text{oct}\left\langle f_j\right\rangle\sin\Omega_e$ yields the simple pendulum of \citepaperI. On the librating side $\left\langle f_\Omega\right\rangle$ does not exist: once per cycle $\mathbf{e}$ crosses $\hat{\mathbf{z}}$ ($e=e_z$, $i_e=0$), where $f_\Omega$ diverges, and the integrated effect of the divergence is not a slow precession but the half-turn (Equation \ref{eq:monodromy}). The azimuth therefore advances by $\pi$ each cycle, $\sin\Omega_e$ alternates in sign, and the kicks \textit{cancel pairwise}, $+\epsilon_\text{oct}\chi\sin\Omega_e,-\epsilon_\text{oct}\chi\sin\Omega_e,\ldots$ - the cancellation observed by \citekdm, who noted that the contribution to $j_z$ vanishes to zeroth order in $j_z$ and accordingly focused on the rotating class. No choice of averaged rate equations can then describe the librating side at first order: the slow dynamics must be sought one order higher, in the correlations between consecutive cycles.

\section{The slow pendulum in the doubled angle}\label{sec:slow_pendulum}

The evolution of $j_z$ ahead organizes into two layers: a fast \textit{zigzag} - the alternate stepping under consecutive, pairwise-cancelling kicks - riding on a \textit{slow layer} that accumulates over many cycles, which this section reduces to a simple pendulum in the doubled angle $2\Omega_e$. Carrying the expansion to the next order (the closed forms quoted here are collected in Appendix \ref{app:coefficients}), the per-cycle map of the slow variables reads
\begin{equation}
\begin{aligned}
    \Omega_{e,n+1}&=\Omega_{e,n}+\pi+F\left(C_K\right)j_z^{\left(n+\frac{1}{2}\right)}-\epsilon_\text{oct}\,c_\delta\left(C_K\right)\sin\Omega_{e,n}\\
    j_z^{\left(n+\frac{3}{2}\right)}&=j_z^{\left(n+\frac{1}{2}\right)}-\epsilon_\text{oct}\,\chi\sin\Omega_{e,n+1}-\epsilon^2_\text{oct}\,\chi_2\sin2\Omega_{e,n+1}
\end{aligned}
\label{eq:map}
\end{equation}
where the integer $n$ counts the $e_\text{max}$ passages as in Section \ref{sec:monodromy}, and half-integer epochs mark the $e_\text{min}$ instants between them - at which the half-turn snaps the azimuth, so that $\Omega_{e,n}$ is the azimuth of $\mathbf{e}$ at the $n$-th $e_\text{max}$ passage and $j_z^{\left(n+\frac{1}{2}\right)}$ is the value of $j_z$ at the $e_\text{min}$ between passages $n$ and $n+1$. Here $F<0$ is the quadrupole phase slip per cycle at small $\left|j_z\right|$,
\begin{equation}
    F\left(C_K\right)=\frac{3}{2k\sqrt{C_z}}\left[2E\left(k\right)-\left(2-k\right)K\left(k\right)\right],
    \label{eq:F}
\end{equation}
with $E\left(m\right)$ the complete elliptic integral of the second kind ($F$ is the librating counterpart of $\left\langle f_\Omega\right\rangle T_e$); $c_\delta>0$ is the first-order octupole phase slip - which, like the kick, alternates sign with each half-turn; and $\chi_2$ is the second-order kick harmonic (Appendix \ref{app:coefficients}). Each line of the map (Equation \ref{eq:map}) is truncated at the lowest order that survives. In the $\Omega_e$ line, $j_z$ is itself $O\left(\epsilon_\text{oct}\right)$ on the slow layer, so the two terms beyond $\pi$ are comparable; in the $j_z$ line, the first-order kick cancels pairwise under the half-turn - it feeds the zigzag, not the slow layer - while $\sin2\Omega_e$ is invariant and the second harmonic survives.

The map carries a period-2 alternation. Two variables absorb it: the parity-corrected phase $\psi$, which strips the exact half-turn advance of Equation \ref{eq:monodromy} and retains only the dynamical slip, and the cleaned $j_z$, denoted $\Jh$, with the zigzag removed:
\begin{equation}
    \psi_n = \Omega_{e,n} + n\pi,\qquad \Jh_n = j_z^{\left(n+\frac{1}{2}\right)} + \frac{\epsilon_\text{oct}\chi}{2}\left(-1\right)^n \sin\psi_n .
    \label{eq:cleaned}
\end{equation}
In these variables the alternating first-order terms cancel over consecutive cycle pairs, and the surviving second-order terms - the correlation of each kick with the phase slip the previous kick produced ($\chi F+c_\delta$) and the second harmonic ($\chi_2$) - drive the continuum limit
\begin{equation}
    \frac{\mathrm{d}\psi}{\mathrm{d}n} = F\,\Jh,
    \qquad
    \frac{\mathrm{d}\Jh}{\mathrm{d}n} = -\frac{\epsilon^2_\text{oct}\Lambda}{4}\sin 2\psi,
    \label{eq:slow_pendulum}
\end{equation}
where
\begin{equation}
    \Lambda\left(C_K\right) = \chi\left(\chi F + c_\delta\right) + 4\chi_2 > 0 .
    \label{eq:Lambda}
\end{equation}
As in \citepaperI-\citepaperIII, the coefficients - all functions of the slowly varying $C_K$ - are evaluated at the initial $C^0_K$. The freezing is not an additional assumption: the secular drift of $C_K$ turns out to be quadratic in $j_z$, hence $O\left(\epsilon^2_\text{oct}\right)$ on the slow trajectories; Section \ref{sec:slaving} derives it in closed form, together with the corrections to this \textit{frozen} model - which become important only at large $\epsilon_\text{oct}$. Under this approximation Equations \ref{eq:slow_pendulum} are again the equations of a simple pendulum - now in the \textit{doubled} angle $2\psi$ - with $\Jh$ as the momentum and $\frac{\mathrm{d}\psi}{\mathrm{d}n}=F\Jh$ as the velocity. The conserved energy is
\begin{equation}
    H=\frac{F}{2}\Jh^2-\frac{\epsilon^2_\text{oct}\Lambda}{8}\cos2\psi .
    \label{eq:H}
\end{equation}
Multiplying by $F$ casts the energy in the velocity form of \citepaperII, $FH=\frac{1}{2}\left(\frac{\mathrm{d}\psi}{\mathrm{d}n}\right)^2+V$ up to an additive constant, with the potential $V=\frac{\epsilon^2_\text{oct}\left|F\right|\Lambda}{8}\left(1+\cos2\psi\right)$, whose minima $\psi=\pm\frac{\pi}{2}$ are the stable fixed points and whose maxima $\psi=0,\pi$ are the unstable ones. We emphasize the structural difference from the rotating side: the coefficients $c_\delta$ and $\chi_2$ are properties of \textit{pairs} of cycles - neither is the cycle average of any static kernel, and neither exists in the rate format of \citekdm.

Conservation of the energy (Equation \ref{eq:H}) resolves the momentum along every slow trajectory by phase:
\begin{equation}
    \Jh^2=\Jh^2_0+\Jh^2_\text{sep}\left(\sin^2\psi-\sin^2\psi_0\right),
    \qquad
    \Jh_\text{sep}\equiv\epsilon_\text{oct}\sqrt{\frac{\Lambda}{2\left|F\right|}} .
    \label{eq:Jsep}
\end{equation}
$\left|\Jh\right|$ peaks at the stable azimuths $\psi=\pm\frac{\pi}{2}$, which every slow trajectory visits, at
\begin{equation}
    \Jh^2_\text{max} = \Jh^2_0 + \Jh^2_\text{sep}\cos^2\psi_0 .
    \label{eq:Jmax}
\end{equation}
Setting $\Jh=0$ in Equation \ref{eq:Jsep} gives $\sin^2\psi=\sin^2\psi_0-\Jh^2_0/\Jh^2_\text{sep}$: a zero crossing exists - the slow pendulum \textit{librates} around $\psi=\pm\frac{\pi}{2}$ - iff $\left|\Jh_0\right|<\Jh_\text{sep}\left|\sin\psi_0\right|$; otherwise $\Jh$ keeps its sign and the pendulum \textit{circulates}. The trajectory through the unstable fixed points, $\left(\psi_0,\Jh_0\right)=\left(0,0\right)$, is the \textit{separatrix} of the slow pendulum; its amplitude $\Jh_\text{sep}$ bounds the excursion of every librating trajectory ($\Jh_\text{max}\leq\Jh_\text{sep}$). The amplitude of the slow $j_z$ oscillation is thus $\propto\epsilon_\text{oct}$ - parametrically smaller than the $\propto\sqrt{\epsilon_\text{oct}}$ amplitude of the rotating side (\citepaperI) - and the slow timescale is $\sim1/\epsilon_\text{oct}$ KLCs rather than $1/\sqrt{\epsilon_\text{oct}}$. Note that the second-order torque does not produce an $\epsilon^{-2}_\text{oct}$ modulation period: the pendulum converts it into a frequency and an amplitude that are both first order - the characteristic period is $2\pi/\left(\epsilon_\text{oct}\sqrt{\left|F\right|\Lambda/2}\right)$ KLCs - sharpening the $t\sim\epsilon^{-2}_\text{oct}t_\text{sec}$ estimate of \citekdm (the large $C_K$-dependent prefactor grows toward the deep librating edge $C_K\to-\frac{3}{2}$, where the cycle degenerates onto the Kozai fixed point and $F$ and $\Lambda$ vanish, and can make the period resemble a steeper power over a limited range of $\epsilon_\text{oct}$).

\begin{figure}
 \begin{centering}
 \includegraphics[width=\columnwidth]{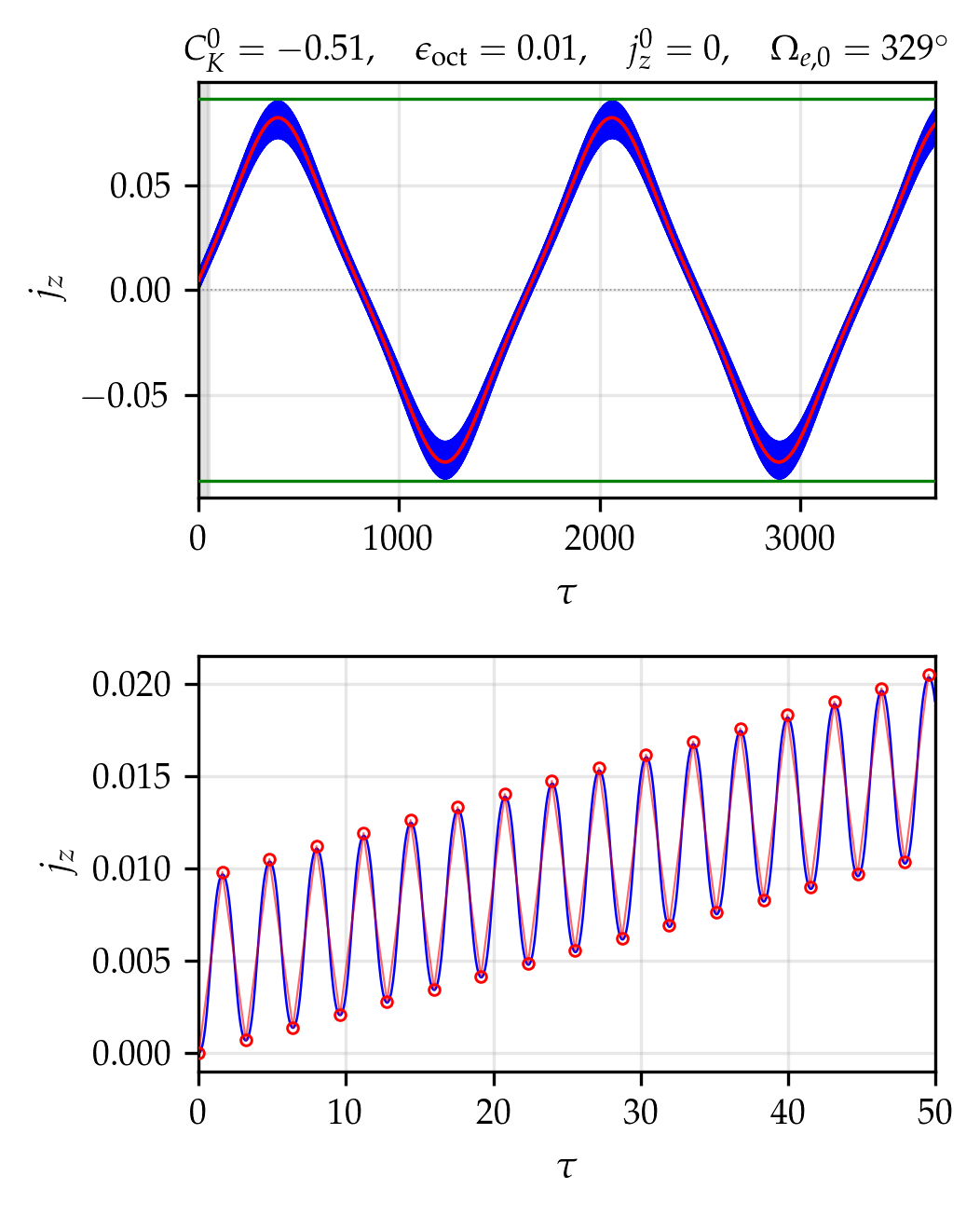}
 \par\end{centering}
 \caption{Results of a numerical integration of the double averaged equations (blue) along with the analytic solution (red), computed from the initial conditions alone. The values of the initial conditions and $\epsilon_\text{oct}$ are shown above the plot. Top panel: $j_z$ as a function of (normalized) time; the red line is the slow momentum $\Jh$ of the doubled-angle pendulum (Equations \ref{eq:slow_pendulum}) and the two green horizontal lines are the predicted extrema of $j_z$, $\pm\left(\Jh_\text{max}+\frac{\epsilon_\text{oct}\chi}{2}\right)$ (Equations \ref{eq:cleaned} and \ref{eq:Jmax}). Bottom panel: zoom on $\tau\leq50$ ($\sim30$ KLCs; shaded in the top panel), resolving the alternating first-order kick (amplitude $\frac{\epsilon_\text{oct}\chi}{2}\left|\sin\psi\right|$, Equation \ref{eq:cleaned}); the red circles are the per-cycle map iterates $j_z^{\left(n+\frac{1}{2}\right)}$ (Equation \ref{eq:map}) at the successive $e_\text{min}$ epochs - one circle per KLC, at each vertex of the zigzag: consecutive kicks cancel pairwise and only the second-order drift accumulates.\label{fig:evolution}}
\end{figure}

Every analytic prediction compared with the numerics in this Letter is computed from the initial conditions alone - no quantity is fitted. An example of a numerical integration of the full DA equations compared with the solution of Equations \ref{eq:slow_pendulum} is shown in Figure \ref{fig:evolution}\footnote{The initial condition of Figure \ref{fig:evolution} has $j^0_z=0$ at the quadrupole $e_\text{min}$ with $\omega_0=+\frac{\pi}{2}$, so $\mathbf{e}\parallel\hat{\mathbf{z}}$ ($e=e_z$) and the instantaneous azimuth of $\mathbf{e}$ is undefined. The quoted $\Omega_{e,0}$ is therefore not an angle of the initial $\mathbf{e}$: it is the azimuth of $\mathbf{e}$ at the first $e_\text{max}$ passage, half a cycle later. It is set by rigidly rotating the initial $\mathbf{j}$ about $\hat{\mathbf{z}}$ - which leaves $\mathbf{e}\parallel\hat{\mathbf{z}}$ unchanged - so that when $\mathbf{e}$ tips away from $\hat{\mathbf{z}}$ toward $e_\text{max}$ its azimuth is the desired value. The same convention is used for every $j^0_z=0$ initial condition in this Letter.}. The fast alternation of amplitude $\frac{\epsilon_\text{oct}\chi}{2}\left|\sin\psi\right|$ (the zigzag, Equation \ref{eq:cleaned}) rides on top of the slow pendulum oscillation of $\Jh$; both are reproduced by the model.

\section{Maximal deviation of $j_z$ and the flip criterion}\label{sec:flip}

\subsection{Maximal deviation of $j_z$ from $j^0_z=0$}\label{subsec:jzmax}

\begin{figure}
 \begin{centering}
 \includegraphics[width=\columnwidth]{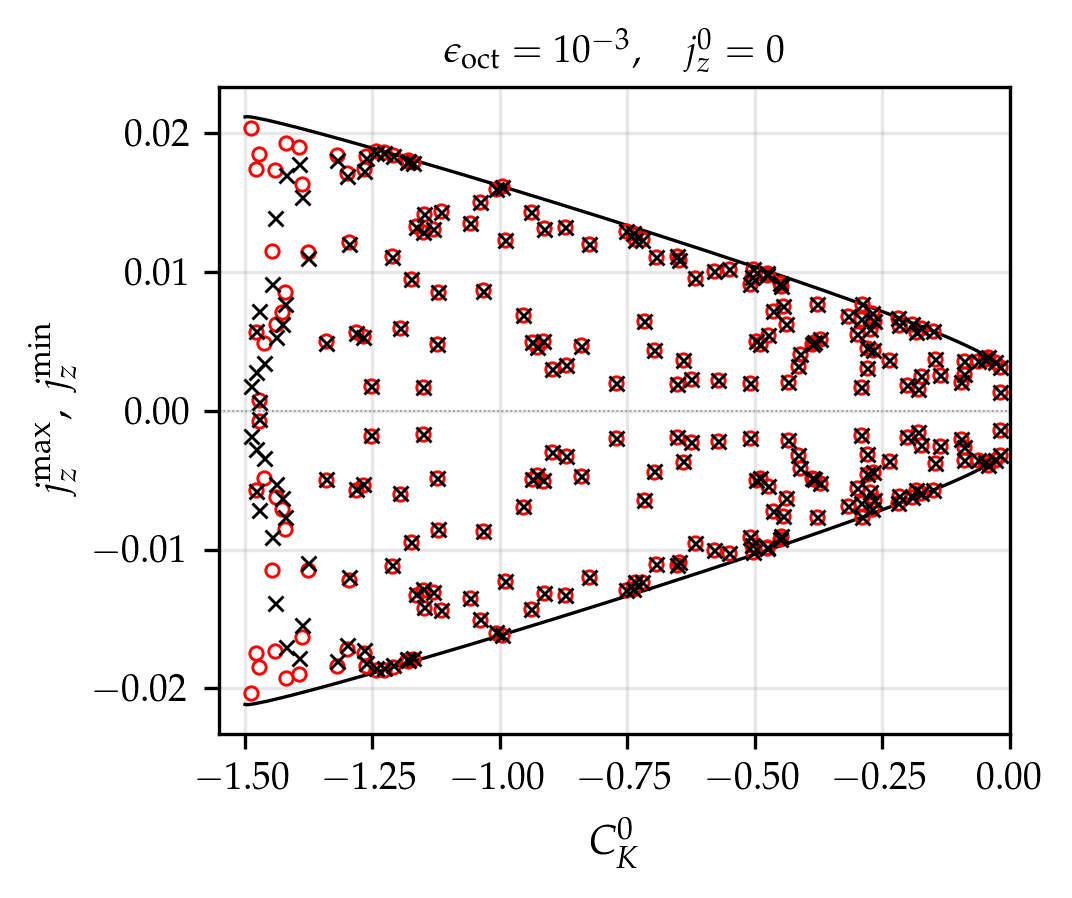}
 \par\end{centering}
 \caption{$j^{\text{max}}_z$ and $j^{\text{min}}_z$ vs. $C^0_K$ when $j^0_z=0$, for $128$ realizations with randomly drawn $\left(C^0_K,\Omega_{e,0}\right)$ and $\epsilon_\text{oct}=10^{-3}$. The results of direct numerical integrations of the double averaged equations are shown using black crosses. The analytic prediction $\pm\left(\Jh_\text{max}+\frac{\epsilon_\text{oct}\chi}{2}\right)$ is marked with open red circles and the envelope (Equation \ref{eq:envelope}) with a black line. Adjacent to the librating edge ($C^0_K\lesssim-1.45$) some crosses fall below their circles: there the excursion is limited by the fixed-point ceiling of Section \ref{sec:slaving} rather than by the slow separatrix.\label{fig:jzminmax}}
\end{figure}

Starting from $j^0_z=0$ at the quadrupole $e_\text{min}$ (analytically known given $\left(C^0_K,j^0_z\right)$ - \citealt{vashkovyak1999}, \citealt{kinoshita2007}, \citealt{antognini2015}, \citealt{basha2025}), the initial condition sits half a cycle before the first $e_\text{max}$ passage - epoch $\left(-\frac{1}{2}\right)$ of the map - so the kick of that first passage sets $j_z^{\left(\frac{1}{2}\right)}=-\epsilon_\text{oct}\chi\sin\psi_0$ (Equation \ref{eq:map}), of which the cleaned momentum keeps half: $\Jh_0=-\frac{\epsilon_\text{oct}\chi}{2}\sin\psi_0$ (Equation \ref{eq:cleaned} at $n=0$). The slow trajectory librates ($\left|\Jh_0\right|<\Jh_\text{sep}\left|\sin\psi_0\right|$, the libration condition of Section \ref{sec:slow_pendulum}, as $\frac{\chi}{2}<\sqrt{\frac{\Lambda}{2\left|F\right|}}$ across the librating window) and $\Jh$ attains $\pm\Jh_\text{max}$ (Equation \ref{eq:Jmax}) at the stable azimuths $\psi=\pm\frac{\pi}{2}$ - exactly where the zigzag reaches its full amplitude - so the predicted extrema of $j_z$ over a slow period are $\pm\left(\Jh_\text{max}+\frac{\epsilon_\text{oct}\chi}{2}\right)$. The maximal and minimal values of $j_z$ obtained in $128$ numerical integrations of the DA equations (Equations \ref{eq:milankovitch_j}-\ref{eq:milankovitch_e}) with $j^0_z=0$ and randomly drawn $\left(C^0_K,\Omega_{e,0}\right)$ - each solved up to $\tau_\text{max}=6\pi T_e/\left(\epsilon_\text{oct}\sqrt{\left|F\right|\Lambda/2}\right)$, three characteristic slow periods (Section \ref{sec:slow_pendulum}) evaluated at $C^0_K$ - are compared with this prediction in Figure \ref{fig:jzminmax} for $\epsilon_\text{oct}=10^{-3}$. The envelope over all azimuths is
\begin{equation}
    j^\text{max}_z\left(C^0_K\right)=\Jh_\text{sep}+\frac{\epsilon_\text{oct}\chi}{2}=\epsilon_\text{oct}\left[\sqrt{\frac{\Lambda}{2\left|F\right|}}+\frac{\chi}{2}\right]
    \label{eq:envelope}
\end{equation}
(slow separatrix plus the fast ripple). Its verification across $\epsilon_\text{oct}=10^{-4}-10^{-1}$ is deferred to Figure \ref{fig:scaling} (Section \ref{sec:slaving}), where the saturation at large $\epsilon_\text{oct}$ is also captured analytically.

\subsection{Flip criterion}\label{subsec:flip_criterion}

\begin{figure}
 \begin{centering}
 \includegraphics[width=\columnwidth]{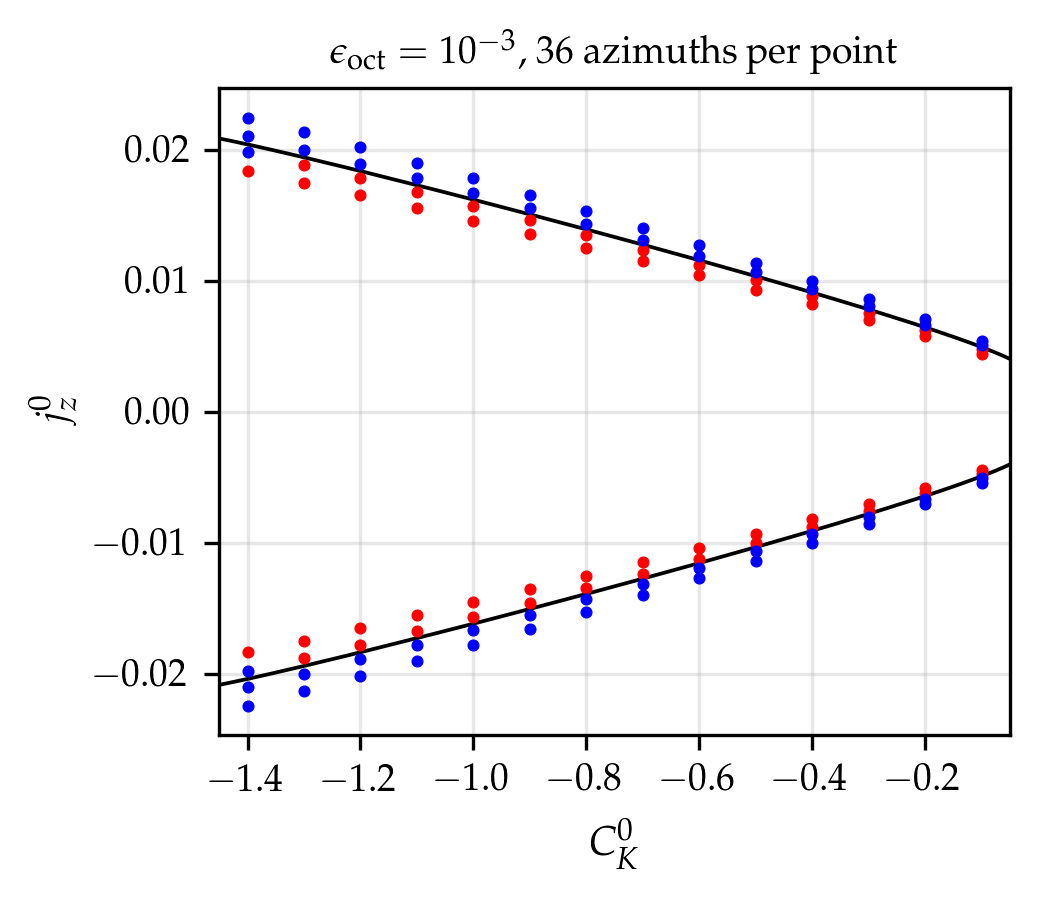}
 \par\end{centering}
 \caption{Parameter space of orbital flips for librating cycles from the solution of the double averaged secular equations, for $\epsilon_\text{oct}=10^{-3}$. Each point represents a set of $36$ numerical integrations with the same $j^0_z$ and $C^0_K$ and a scan of initial azimuths; the values $j^0_z=\pm\left\{0.9,0.97,1.03,1.1\right\}\times j_\text{crit}$ deliberately bracket the predicted boundary on both sides of $j_z=0$. Red points mark integrations where $j_z$ crossed zero for some azimuth and blue points those where $j_z$ kept its sign for all. The black lines are the analytic boundary $\pm j_\text{crit}$ (Equation \ref{eq:jcrit}).\label{fig:flipmap}}
\end{figure}

A flip - zero crossing of $j_z$ - is equivalent to a zero crossing of the slow momentum $\Jh$, which occurs only on librating slow trajectories (Section \ref{sec:slow_pendulum}). The fast ripple alone cannot flip a circulating trajectory: along it $\Jh^2-\Jh^2_\text{sep}\sin^2\psi$ is a positive constant (Equation \ref{eq:Jsep}), so $\left|\Jh\right|\geq\Jh_\text{sep}\left|\sin\psi\right|$ at every phase, while the ripple carries the same $\left|\sin\psi\right|$ with a smaller coefficient: $\frac{\chi}{2}<\sqrt{\frac{\Lambda}{2\left|F\right|}}$ across the entire librating window. Given $\epsilon_\text{oct}$ and initial values $C^0_K,j^0_z,\psi_0$ (at the quadrupole $e_\text{min}$), and abbreviating the kick offset $b=\frac{\epsilon_\text{oct}\chi}{2}$ (so that $\Jh_0=j^0_z-b\sin\psi_0$; Section \ref{subsec:jzmax}), the criterion is the libration condition at the initial data:
\begin{equation}
    \text{flip}\iff\left|j^0_z-b\sin\psi_0\right|<\Jh_\text{sep}\left|\sin\psi_0\right|.
    \label{eq:flip_criterion}
\end{equation}
Maximizing over $\psi_0$, a flip is possible for some initial azimuth iff $\left|j^0_z\right|<j_\text{crit}$ with
\begin{equation}
    j_\text{crit}\left(C^0_K,\epsilon_\text{oct}\right)=\Jh_\text{sep}+b=\epsilon_\text{oct}\left[\sqrt{\frac{\Lambda}{2\left|F\right|}}+\frac{\chi}{2}\right],
    \label{eq:jcrit}
\end{equation}
identical to the envelope (Equation \ref{eq:envelope}): the maximal excursion of $j_z$ starting from $j^0_z=0$ (Section \ref{subsec:jzmax}, Figure \ref{fig:jzminmax}) is also the maximal $\left|j^0_z\right|$ from which a flip is possible - as on the rotating side ($C^0_K>0$), where a single line bounds both the flip map and the excursions from $j^0_z=0$ (Figures 2 and 3 of \citepaperI). A comparison between a numerical flip map and the analytic boundary as a function of $j^0_z$ and $C^0_K$ is shown in Figure \ref{fig:flipmap} for $\epsilon_\text{oct}=10^{-3}$: each point stands for $36$ numerical integrations with initial azimuths equally spaced by $10^\circ$ (each up to the $\tau_\text{max}$ of Section \ref{subsec:jzmax}) - red if $j_z$ crossed zero for some azimuth, blue if it kept its sign for all - with $j^0_z$ at fixed multiples of $j_\text{crit}$ (Equation \ref{eq:jcrit}), just inside and just outside the predicted boundary, for both signs of $j^0_z$; the black lines, $\pm j_\text{crit}$, trace the border between the red and the blue points across the librating window.

\begin{figure}
 \begin{centering}
 \includegraphics[width=\columnwidth]{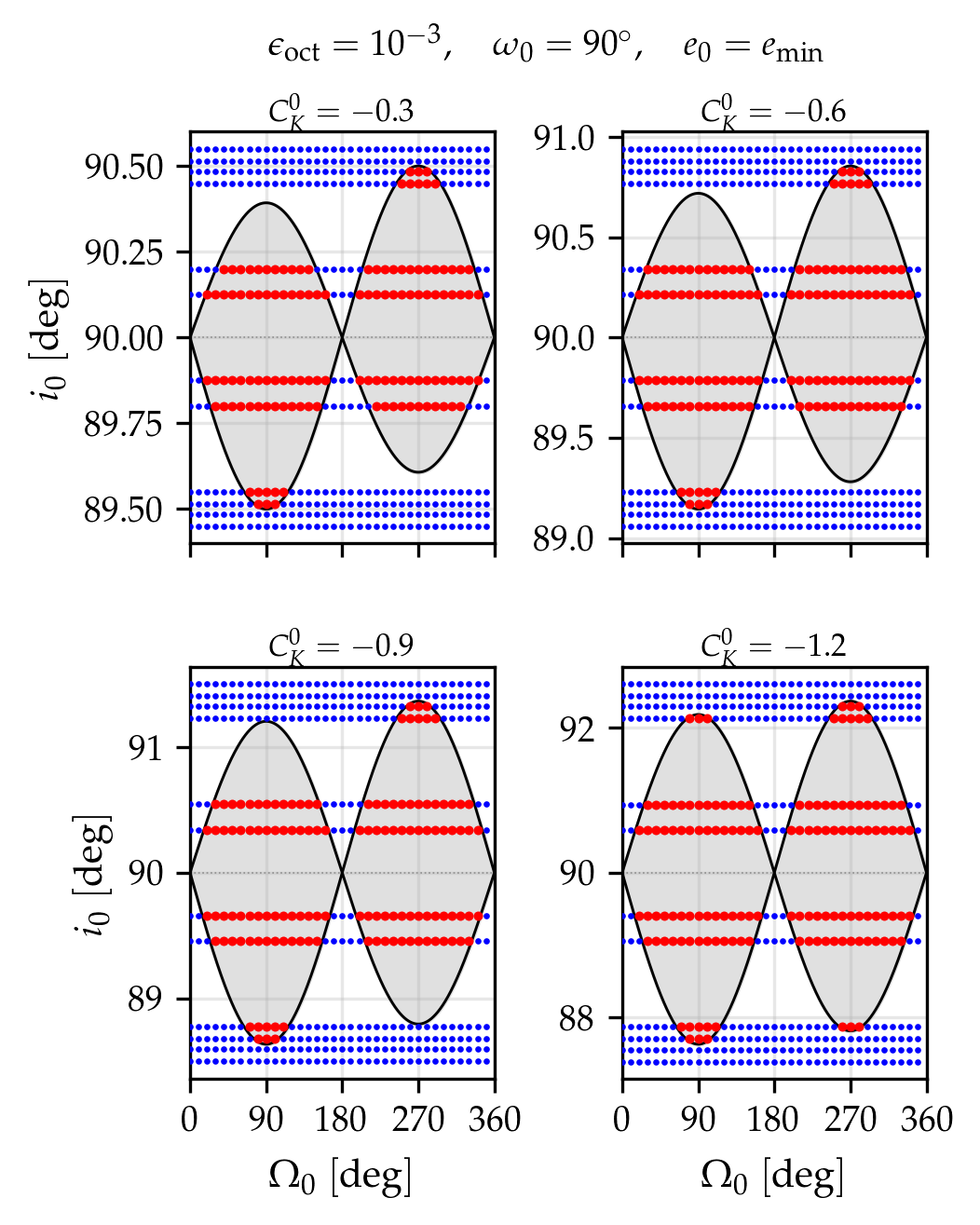}
 \par\end{centering}
 \caption{The flip window in the observable $\left(\Omega_0,i_0\right)$ plane for $\epsilon_\text{oct}=10^{-3}$, one $C^0_K$ per panel; initial conditions at the quadrupole $e_\text{min}$ with $\omega_0=90^\circ$ (Section \ref{subsec:flip_criterion}). Points are numerical integrations scanning the initial azimuth at fixed $j^0_z$: rows bracketing the boundary at $j^0_z=\pm\left\{0.9,0.97,1.03,1.1\right\}\times j_\text{crit}$ and interior rows at $j^0_z=\pm\left\{0.25,0.4\right\}\times j_\text{crit}$; red if $j_z$ crossed zero, blue if it kept its sign. The grey region bounded by the black curves is the analytic criterion (Equation \ref{eq:flip_criterion}): the tips of the lens reach $\left|j^0_z\right|=j_\text{crit}$ (Equation \ref{eq:jcrit}) and the asymmetry between its two lobes is the kick offset $b=\frac{\epsilon_\text{oct}\chi}{2}$.\label{fig:flipplane}}
\end{figure}

\begin{figure}
 \begin{centering}
 \includegraphics[width=\columnwidth]{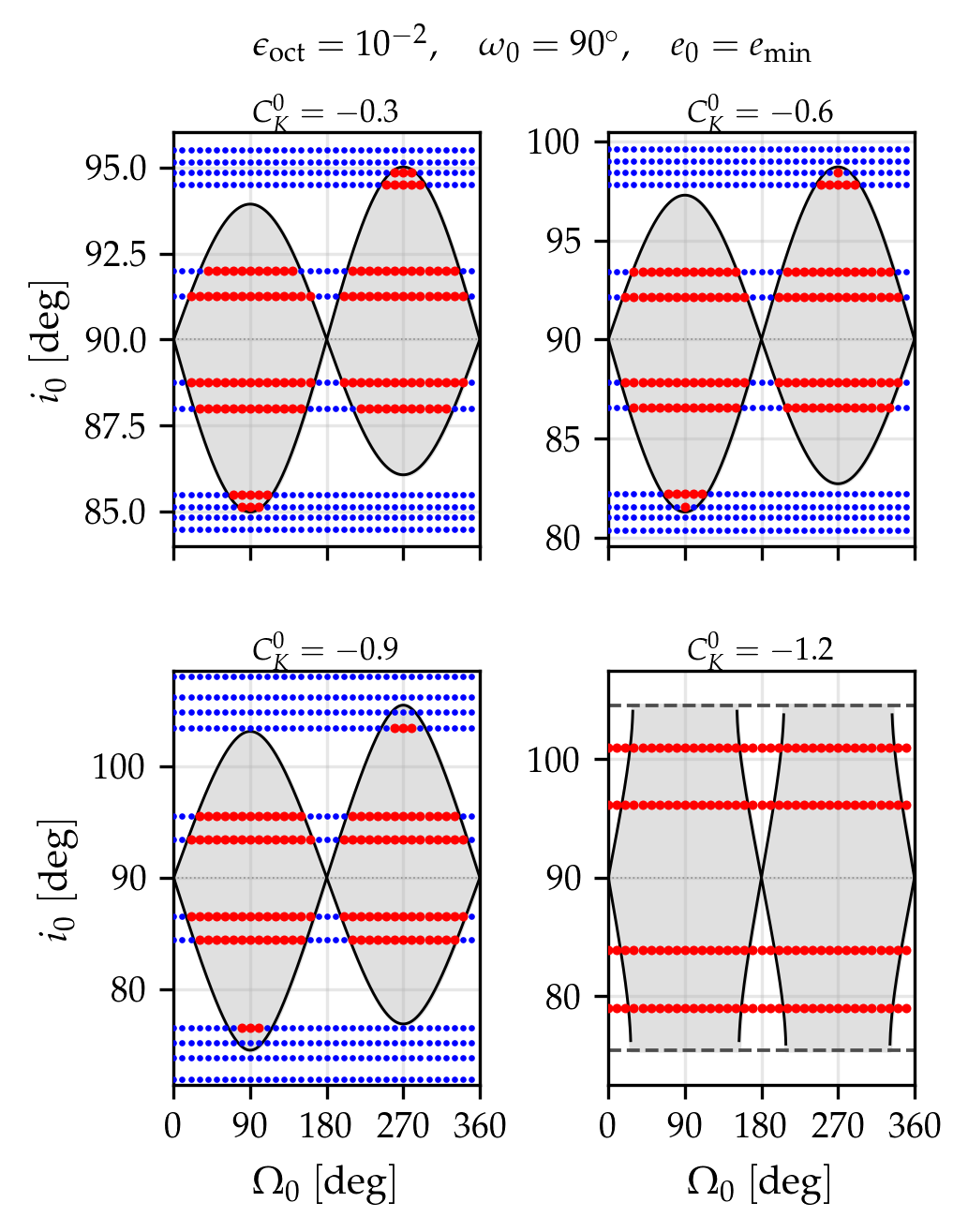}
 \par\end{centering}
 \caption{Same as Figure \ref{fig:flipplane} for $\epsilon_\text{oct}=10^{-2}$ - the same rows in units of the ten times larger $j_\text{crit}$, so the window is ten times wider (at $C^0_K=-0.9$ it spans $i_0\approx75^\circ-105^\circ$). Near the boundary the measured flips retreat into the tips of the lens: at $C^0_K=-0.9$ the $\pm0.97\,j_\text{crit}$ rows no longer flip at any azimuth. At $C^0_K=-1.2$ the predicted window outgrows the librating class itself: librating states exist only for $\left|j^0_z\right|\leq\frac{1}{5}\left(\sqrt{15}-\sqrt{-10\,C^0_K}\right)$ (inverting the fixed-point floor on $C_K$, Equation \ref{eq:floor}; dashed lines), so the bracketing rows do not exist and the criterion boundary survives only near the pinches; the interior rows do exist, and every one of their $144$ integrations flipped, \textit{including at the pinches where Equation \ref{eq:flip_criterion} forbids it}: the frozen lens has dissolved (Section \ref{sec:slaving}, Figure \ref{fig:flipfraction}).\label{fig:flipplane_deep}}
\end{figure}

Figures \ref{fig:flipplane} and \ref{fig:flipplane_deep} recast the flip criterion (Equation \ref{eq:flip_criterion}) in the plane of two observable orbital elements: the initial inclination $i_0$ and the initial longitude of the ascending node $\Omega_0$, measured (like $\Omega_e$) from the outer pericenter. The node is an angle distinct from the azimuth $\Omega_e$, but for the initial conditions of this section it supplies the phase of the criterion: $\psi_0=\Omega_0$ up to a sub-degree quadrupole offset. Equation \ref{eq:flip_criterion} then bounds a lens-shaped window around $i_0=90^\circ$ whose two lobes are rendered asymmetric by the kick offset $b$: the deepest prograde ($i_0<90^\circ$) reach sits at $\Omega_0=90^\circ$, the deepest retrograde ($i_0>90^\circ$) reach at $\Omega_0=270^\circ$, and the lobes meet - the window closes - at the \textit{pinches}, $\Omega_0=0^\circ,180^\circ$. Both figures overlay rows of numerical integrations scanning the initial azimuth at fixed $j^0_z$: rows bracketing the boundary at $j^0_z=\pm\left\{0.9,0.97,1.03,1.1\right\}\times j_\text{crit}$ and \textit{interior} rows, well inside the window, at $j^0_z=\pm\left\{0.25,0.4\right\}\times j_\text{crit}$; red marks integrations in which $j_z$ crossed zero, blue those in which it kept its sign. In Figure \ref{fig:flipplane} ($\epsilon_\text{oct}=10^{-3}$) the black curves trace exactly the border between the red and the blue points: along the interior rows the flips switch on through each lobe and off around the pinches; the rows just inside the boundary flip only adjacent to the tips and those just outside at no azimuth; and the two lobes fill asymmetrically, following the offset $b$. At the ten times larger $\epsilon_\text{oct}$ of Figure \ref{fig:flipplane_deep} the two shallow panels ($C^0_K=-0.3,-0.6$) show the same agreement; at $C^0_K=-0.9$ the boundary is slightly off - the flips of the rows just inside retreat into the tips - and at $C^0_K=-1.2$ it fails altogether: every integration flipped, including at the pinches, where Equation \ref{eq:flip_criterion} forbids a flip. Section \ref{sec:slaving} traces this to the fixed-point ceiling and locates the transition.

\section{The fixed-point ceiling and the energy constraint on $C_K$}\label{sec:slaving}

\begin{figure}
 \begin{centering}
 \includegraphics[width=\columnwidth]{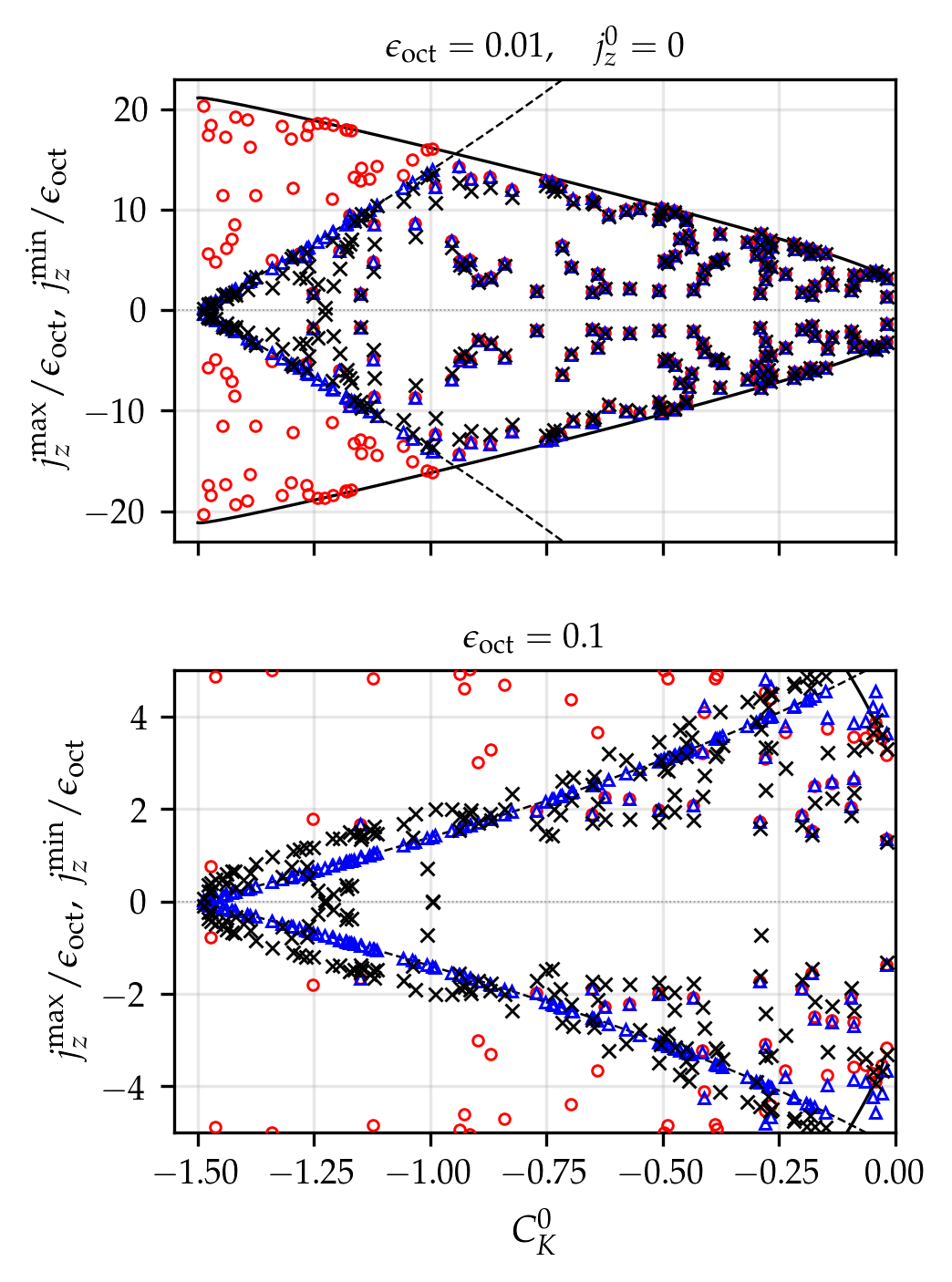}
 \par\end{centering}
 \caption{$j^{\text{max}}_z$ and $j^{\text{min}}_z$ in units of $\epsilon_\text{oct}$ vs. $C^0_K$ when $j^0_z=0$, for the $128$ realizations of Figure \ref{fig:jzminmax}, at $\epsilon_\text{oct}=10^{-2}$ (top) and $\epsilon_\text{oct}=10^{-1}$ (bottom; note the zoomed vertical range). Black crosses: numerical DA extrema. Red open circles: the frozen-$C_K$ prediction $\pm\left(\Jh_\text{max}+\frac{\epsilon_\text{oct}\chi}{2}\right)$ - exactly linear in $\epsilon_\text{oct}$, so in these units it is the same pattern in both panels and, rescaled, the pattern of Figure \ref{fig:jzminmax} - bounded by the envelope (Equation \ref{eq:envelope}, solid black). Blue triangles: the prediction including the energy constraint and the fixed-point ceiling (Equations \ref{eq:slaving}-\ref{eq:ceiling}). The black dashed line is the bare ceiling $\pm\Jh_\text{ceil}/\epsilon_\text{oct}$ (Equation \ref{eq:ceiling}), a bound on the slow momentum $\Jh$: its intersection with the envelope splits the window - to its right the frozen pendulum is exact, to its left the slow momentum saturates at the ceiling (the capped prediction collapses onto it there, while the measured $j_z$ retains part of the fast ripple $\frac{\epsilon_\text{oct}\chi}{2}\left|\sin\psi\right|$, Equation \ref{eq:cleaned}). At $\epsilon_\text{oct}=10^{-2}$ the ceiling bites only at $C^0_K\lesssim-1$; at $10^{-1}$ it rules the entire window.\label{fig:jzminmax_deep}}
\end{figure}

\begin{figure}
 \begin{centering}
 \includegraphics[width=\columnwidth]{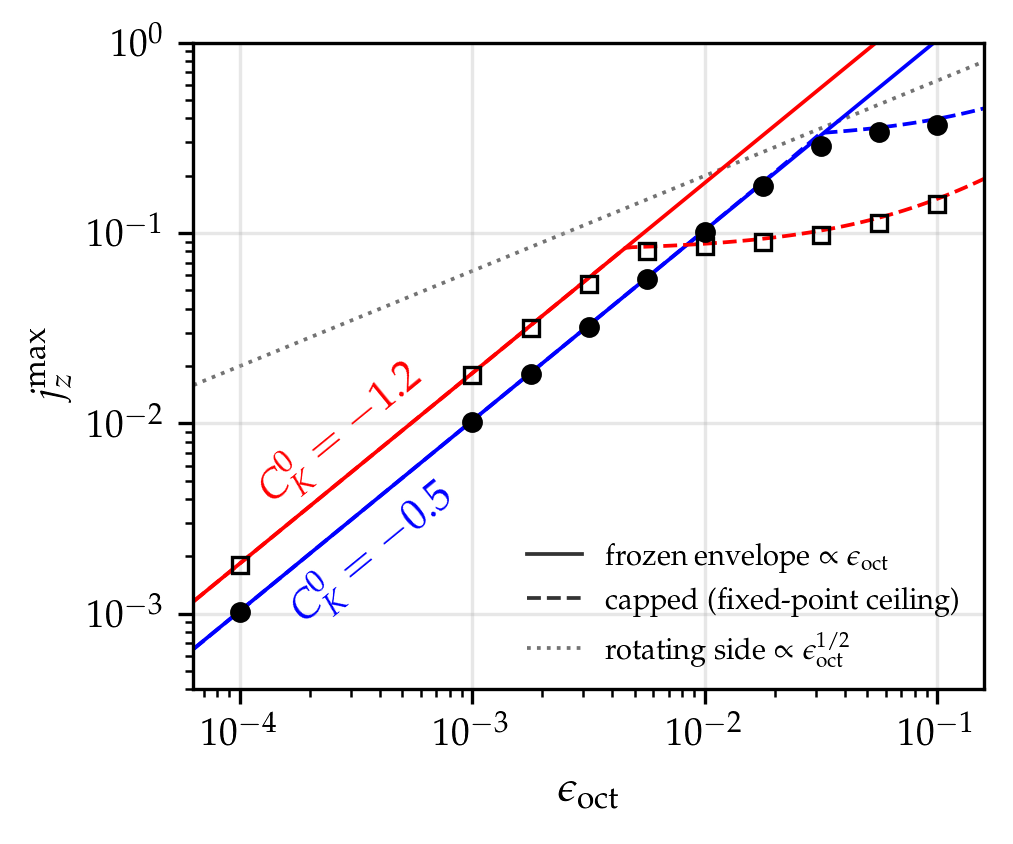}
 \par\end{centering}
 \caption{The maximal attainable value of $j^{\text{max}}_z$ (over initial azimuths) vs. $\epsilon_{\text{oct}}$ for initial conditions with $j^0_z=0$, at two values of $C^0_K$ (blue lines: $C^0_K=-0.5$; red lines: $C^0_K=-1.2$). Black symbols: numerical solutions of the double averaged equations - for each $\epsilon_\text{oct}$, the maximum over $36$ initial azimuths equally spaced by $10^{\circ}$ at exactly $C^0_K$ (filled circles for $C^0_K=-0.5$, open squares for $C^0_K=-1.2$). Solid lines: the frozen-$C_K$ envelope (Equation \ref{eq:envelope}), $\propto \epsilon_{\text{oct}}$. Dashed lines: the all-azimuth envelope of the model including the energy constraint and the fixed-point ceiling (Equations \ref{eq:slaving}-\ref{eq:ceiling}) at exactly $C^0_K$ - it coincides with the frozen envelope at small $\epsilon_\text{oct}$, departs from it as the ceiling is approached, and saturates at $\Jh_\text{ceil}$ plus the fast ripple $\frac{\epsilon_\text{oct}\chi}{2}$ (the maxima over the discrete azimuth grid sit at or just below it). The dotted grey guide shows the $\propto\sqrt{\epsilon_\text{oct}}$ scaling of the rotating side for contrast, anchored at $j^\text{max}_z=2\times10^{-2}$ at $\epsilon_\text{oct}=10^{-4}$ (the $C_K=0.6$ curve of Figure 4 of \citepaperI); it crosses the frozen envelopes at $\epsilon_\text{oct}$ of a few $\times10^{-2}$ (Section \ref{sec:discussion}).\label{fig:scaling}}
\end{figure}

The frozen-$C_K$ model of Section \ref{sec:slow_pendulum} evaluates $\chi,F,\Lambda$ at $C^0_K$ and carries no knowledge of the boundary of the phase space. The boundary is exact and purely geometric: minimizing $C_K$ over all states at fixed $j_z$ gives the floor
\begin{equation}
    C_K\geq-\frac{3}{2}+\sqrt{15}\left|j_z\right|-\frac{5}{2}j^2_z,
    \label{eq:floor}
\end{equation}
with equality on the Kozai fixed-point family, where the cycle degenerates to a point - so already at the frozen $C^0_K$ the floor caps the attainable $\left|j_z\right|$ (the dashed lines in the $C^0_K=-1.2$ panel of Figure \ref{fig:flipplane_deep}). Along the actual flow the cap sits slightly lower. The flow of Equations \ref{eq:milankovitch_j}-\ref{eq:milankovitch_e} conserves the time-independent potential $\phi=\phi_\text{quad}+\epsilon_\text{oct}\phi_\text{oct}$ exactly, and solving Equation \ref{eq:CK} for $\phi_\text{quad}$ turns this into $C_K+\frac{1}{2}j^2_z=\text{const}-\frac{4}{3}\epsilon_\text{oct}\phi_\text{oct}$. Dropping the octupole piece to obtain a secular relation between $\bar{C}_K$ and $\Jh$ is the same step as on the rotating side (\citekdm): naively it looks unjustified, because $\epsilon_\text{oct}\phi_\text{oct}$ is $O\left(\epsilon_\text{oct}\right)$ while $\frac{1}{2}j^2_z$ is only $O\left(\epsilon_\text{oct}\right)$ for rotating cycles and $O\left(\epsilon^2_\text{oct}\right)$ here. What matters is not that instantaneous ordering but that the octupole dressing does not accumulate secularly - on the rotating side because $\phi_\text{oct}$ is small through most of each KLC (\citepaperI); here, in addition, its cycle mean alternates under the half-turn like every other first-order effect. The secular parts (denoted by an overbar) then obey the \textit{energy constraint}
\begin{equation}
    \bar{C}_K\left(n\right)=C^0_K-\frac{1}{2}\left(\Jh^2_n-\Jh^2_0\right)
    \label{eq:slaving}
\end{equation}
up to a bounded, non-accumulating $O\left(\epsilon_\text{oct}\right)$ remainder. As the pendulum grows $\left|\Jh\right|$, Equation \ref{eq:slaving} lowers $\bar{C}_K$ while the floor at $\left|j_z\right|=\left|\Jh\right|$ rises, and the two make contact at a finite momentum. Eliminating $\bar{C}_K$ between Equations \ref{eq:floor} and \ref{eq:slaving} gives, with $\tilde{\Delta}=C^0_K+\frac{3}{2}+\frac{1}{2}\Jh^2_0$,
\begin{equation}
    \Jh_\text{ceil}=\frac{\sqrt{15}-\sqrt{15-8\tilde{\Delta}}}{4},
    \label{eq:ceiling}
\end{equation}
and the attainable excursion saturates at $\min\left(\Jh_\text{max},\Jh_\text{ceil}\right)$. This \textit{fixed-point ceiling} is non-perturbative (independent of $\epsilon_\text{oct}$) and undercuts the frozen prediction for all $C^0_K\lesssim-1.1$ at $\epsilon_\text{oct}=10^{-2}$, and almost everywhere at $\epsilon_\text{oct}=10^{-1}$. Figure \ref{fig:jzminmax_deep} repeats Figure \ref{fig:jzminmax} - the extrema of $j_z$ attained by the same $128$ realizations - at these two values of $\epsilon_\text{oct}$, with the vertical axis in units of $\epsilon_\text{oct}$: in these units the frozen prediction (red circles) is the same pattern in both panels, while the bare ceiling $\pm\Jh_\text{ceil}/\epsilon_\text{oct}$ (black dashed) moves down as $\epsilon_\text{oct}$ grows; wherever it undercuts the frozen circles, the measured extrema (black crosses) abandon them for the capped prediction (blue triangles). Figure \ref{fig:scaling} summarizes the full ladder, $\epsilon_\text{oct}=10^{-4}-10^{-1}$: the maximal $j^\text{max}_z$ over initial azimuths at $C^0_K=-0.5$ and $-1.2$ (black symbols) is compared with the frozen envelope (solid lines, colored by $C^0_K$, Equation \ref{eq:envelope}) and with the all-azimuth envelope of the capped model (dashed lines, same colors) - the measured maxima follow the frozen envelope, $\propto\epsilon_\text{oct}$, until the ceiling intervenes ($\epsilon_\text{oct}\approx5\times10^{-3}$ at $C^0_K=-1.2$, $\approx3\times10^{-2}$ at $-0.5$) and saturate beyond it, exactly as the capped pendulum predicts.

\begin{figure}
 \begin{centering}
 \includegraphics[width=\columnwidth]{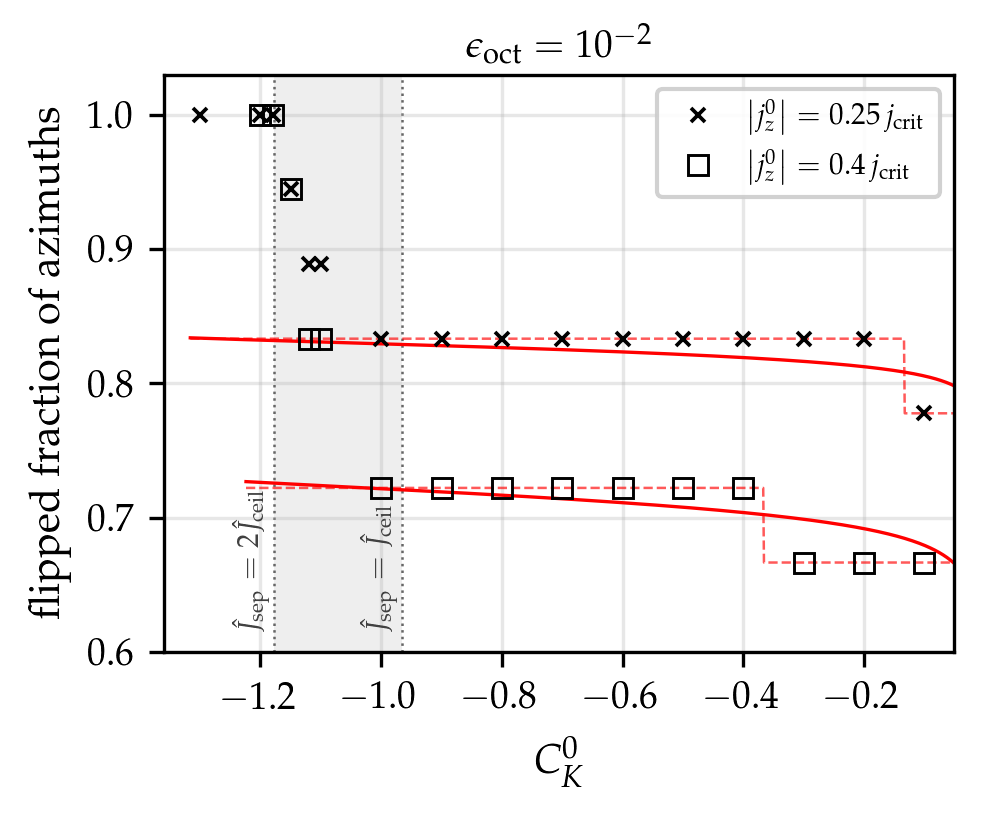}
 \par\end{centering}
 \caption{The dissolution of the flip lens under the fixed-point ceiling: the fraction of initial azimuths that flip vs. $C^0_K$ at $\epsilon_\text{oct}=10^{-2}$, along the interior rows $j^0_z=\pm\left\{0.25,0.4\right\}\times j_\text{crit}$ of Figure \ref{fig:flipplane_deep} ($36$ azimuths per row, mirrored rows combined; crosses $\left|j^0_z\right|=0.25\,j_\text{crit}$, open squares $0.4\,j_\text{crit}$). Solid red: the frozen-lens fraction $P_\text{flip}$ (Equation \ref{eq:flip_fraction}); dashed red: the same two arcs counted on the $10^\circ$ azimuth grid of the data - the exact prediction for these ensembles, deviating from the continuum fraction by at most one azimuth per arc in either direction. The dotted verticals, $\Jh_\text{sep}=\Jh_\text{ceil}$ and $\Jh_\text{sep}=2\,\Jh_\text{ceil}$ (Equations \ref{eq:Jsep} and \ref{eq:ceiling}), bound the shaded band: to its right the measured fractions sit on the gridded prediction point by point, inside it they break upward from $C^0_K\approx-1.1$ as the ceiling captures azimuths outside the lens, and at $\Jh_\text{sep}=2\,\Jh_\text{ceil}$ they saturate at unity - the lens has dissolved.\label{fig:flipfraction}}
\end{figure}

The same ceiling dissolves the flip lens (the $C^0_K=-1.2$ panel of Figure \ref{fig:flipplane_deep}). The quantity to follow is the flip \textit{fraction} of an ensemble uniform in the initial azimuth: at fixed $j^0_z$ the frozen criterion (Equation \ref{eq:flip_criterion}) is met on two arcs of the azimuth circle, one through each lobe of the lens ($\sigma\equiv\left|\sin\psi_0\right|$, $r\equiv\left|j^0_z\right|/j_\text{crit}$) - $\sigma>r$ on the half-circle where $\sin\psi_0$ carries the sign of $j^0_z$, crossing the lobe that the kick offset $b$ deepens, and $\sigma>r\frac{\Jh_\text{sep}+b}{\Jh_\text{sep}-b}$ on the opposite half, crossing the lobe it shallows - of total fraction
\begin{equation}
\begin{aligned}
    P_\text{flip}\left(r\right)=\frac{1}{2\pi}\Big[&\left(\pi-2\arcsin r\right)_+\\
    &+\Big(\pi-2\arcsin\min\Big(1,\frac{\Jh_\text{sep}+b}{\Jh_\text{sep}-b}\,r\Big)\Big)_+\Big],
\end{aligned}
    \label{eq:flip_fraction}
\end{equation}
where $\left(x\right)_+=\max\left(x,0\right)$, interpolating between $P_\text{flip}\left(0\right)=1$ and a square-root closing at the tips, $P_\text{flip}\simeq\frac{1}{\pi}\sqrt{2\left(1-r\right)}$ as $r\to1$. The criterion competes with the ceiling through the ratio $\Jh_\text{sep}/\Jh_\text{ceil}$, which grows $\propto\epsilon_\text{oct}$ and toward the deep edge of the librating window, $C^0_K\to-\frac{3}{2}$, where the ceiling collapses linearly ($\Jh_\text{ceil}\simeq\frac{\tilde{\Delta}}{\sqrt{15}}$, Equation \ref{eq:ceiling}) while $\Jh_\text{sep}$ remains finite. Once the ratio exceeds unity the ceiling enters the flip map. A circulating trajectory peaks above $\Jh_\text{sep}$ (Equation \ref{eq:Jmax}) at the stable azimuths, which it visits within every slow period: with $\Jh_\text{sep}>\Jh_\text{ceil}$, every circulating trajectory in the strip where librating states exist (inverting Equation \ref{eq:floor}; the dashed lines of Figure \ref{fig:flipplane_deep}) makes contact with the ceiling within its first slow period - including at azimuths for which Equation \ref{eq:flip_criterion} predicts no flip. Contact - the secular $\bar{C}_K$ (Equation \ref{eq:slaving}) meeting the floor (Equation \ref{eq:floor}), the cycle degenerating onto the fixed-point family - does not by itself reverse $j_z$: it pins $\left|\Jh\right|$ at $\Jh_\text{ceil}$ while the slow phase keeps advancing, and the trajectory leaves the ceiling, at some phase $\psi_w$, on a new contour: Equation \ref{eq:Jsep} with $\left(\psi_0,\Jh_0\right)$ replaced by $\left(\psi_w,\Jh_\text{ceil}\right)$. The new contour is librating - $j_z$ crosses zero - iff $\Jh_\text{ceil}<\Jh_\text{sep}\left|\sin\psi_w\right|$ (the libration condition of Section \ref{sec:slow_pendulum} at the new data): $j_z$ keeps its sign only when the departure happens near a pinch azimuth, $\left|\sin\psi_w\right|<\Jh_\text{ceil}/\Jh_\text{sep}$. These no-flip departures occupy two arcs around $\psi_w=0,\pi$, a fraction $\frac{2}{\pi}\arcsin\left(\Jh_\text{ceil}/\Jh_\text{sep}\right)$ of the circle: the whole circle when $\Jh_\text{sep}=\Jh_\text{ceil}$, shrinking to a third of it ($\frac{2}{\pi}\arcsin\frac{1}{2}=\frac{1}{3}$) when $\Jh_\text{sep}$ reaches $2\,\Jh_\text{ceil}$. Figure \ref{fig:flipfraction} shows the dissolution directly: for each $C^0_K$, the flipped fraction of the azimuth grid along the interior rows of Figure \ref{fig:flipplane_deep} ($\left|j^0_z\right|=\left\{0.25,0.4\right\}\times j_\text{crit}$), compared with the frozen prediction - the two arcs of Equation \ref{eq:flip_fraction}, drawn both as their continuum fraction and as their count on the $10^\circ$ grid of the ensemble. The sequence above is traced exactly: through $\Jh_\text{sep}=\Jh_\text{ceil}$ (at $C^0_K=-0.97$) the measured fractions sit on the gridded prediction point by point; excess flips accumulate from $C^0_K\approx-1.1$, the surviving azimuths retreating into the pinches; and at $\Jh_\text{sep}\approx2\,\Jh_\text{ceil}$ ($C^0_K=-1.18$) the lens dissolves - every existing librating state flips, the regime of the deepest panel of Figure \ref{fig:flipplane_deep}.

Beyond setting the ceiling (Equation \ref{eq:ceiling}), the energy constraint is an accounting statement: the secular wander of $C_K$ is not a new degree of freedom - it is determined, through Equation \ref{eq:slaving}, by the momentum the pendulum already evolves. Since $\Jh=O\left(\epsilon_\text{oct}\right)$ the wander is $O\left(\epsilon^2_\text{oct}\right)$: the frozen-$C_K$ model is essentially exact for $\epsilon_\text{oct}\lesssim10^{-3}$, and even at $\epsilon_\text{oct}=10^{-1}$ the drift of the coefficients along $\bar{C}_K\left(\Jh\right)$ shifts the predictions only at the per-cent level - the corrections that move the figures are the ceiling's.

\section{Discussion}\label{sec:discussion}

This Letter completes, on the librating side, the reduction of the high-eccentricity octupole dynamics to simple pendulums. The two sides of the $C_K$ phase space fork at a single geometric fact and diverge from there. On the rotating side the azimuth $\Omega_e$ is a slow, smoothly accumulating phase; the octupole drive is a rate; the slow model is a pendulum in $\Omega_e$ with torque $O\left(\epsilon_\text{oct}\right)$; excursions scale as $\sqrt{\epsilon_\text{oct}}$; and flips are delineated by the separatrix of that pendulum (\citekdm, \citepaperI). On the librating side the azimuth snaps by $\pi$ every cycle (Equation \ref{eq:monodromy}); the same transfer kernel $f_j$ - with its cycle average continuing analytically across $C_K=0$ (Equation \ref{eq:fj_librating}) - is demoted from a rate to an alternating kick that cancels pairwise; the slow model is a pendulum in the \textit{doubled} angle with torque $O\left(\epsilon^2_\text{oct}\right)$ built from two-cycle correlations ($c_\delta$, $\chi_2$) that have no rate-format analogue; excursions scale as $\epsilon_\text{oct}$; and flips occupy an explicit window around $\psi_0=\pm\frac{\pi}{2}$ (Equation \ref{eq:flip_criterion}). The rotating-side resonance at $C_K\approx0.112$ (where $\left\langle f_\Omega\right\rangle=0$) has no librating counterpart: $F<0$ throughout the window (Equation \ref{eq:F}), and the librating kick never vanishes (Section \ref{sec:kick}).

The two scalings should not be read as a practical hierarchy. Asymptotically it is real - as $\epsilon_\text{oct}\to0$ the rotating $\sqrt{\epsilon_\text{oct}}$ dwarfs the librating $\epsilon_\text{oct}$, the content of the leading-order cancellation of \citekdm, and the reading implicit in the lineage that followed - but the prefactors lean the other way. The librating envelope (Equation \ref{eq:envelope}) carries $\frac{\chi}{2}+\sqrt{\frac{\Lambda}{2\left|F\right|}}\approx10$ at $C^0_K=-0.5$ ($\approx18$ at $-1.2$), an order of magnitude above unity, while the rotating-side numerical maxima of \citepaperI (the black curves of Figure 4 there, at $C_K=0.05$ and $0.6$) fall below their $\sqrt{\epsilon_\text{oct}}$ analytic tracks at large $\epsilon_\text{oct}$ and sit near $j^\text{max}_z\sim0.2-0.3$ for $\epsilon_\text{oct}\sim10^{-2}-10^{-1}$. Figure \ref{fig:scaling} shows it directly: at $\epsilon_\text{oct}=3\times10^{-2}$ the measured maxima at $C^0_K=-0.5$ reach $j^\text{max}_z=0.29$, already above those rotating numerical values, before the fixed-point ceiling caps them near $j^\text{max}_z\approx0.37$; at $C^0_K=-1.2$ the ceiling is lower, and holds the curve near $j^\text{max}_z\approx0.1$. At octupole strengths routine among hierarchical triples, $\epsilon_\text{oct}\sim10^{-2}-10^{-1}$, the librating class thus delivers $j_z$ excursions comparable to - and in part of the window exceeding - those of the rotating class - what separates the two sides there is not magnitude but structure: the amplitude scales as $\epsilon_\text{oct}$ rather than $\epsilon_\text{oct}^{1/2}$, the slow timescale as $\sim\epsilon_\text{oct}^{-1}$ rather than $\sim\epsilon_\text{oct}^{-1/2}$ KLCs, and the flip anatomy itself (azimuth windows, Equation \ref{eq:flip_criterion}, and their ceiling-driven dissolution).

It is instructive to contrast this with \citepaperIII, which treats a different dynamical system - a constantly precessing quadrupole - and where librating KLCs \textit{are} treated successfully with rate-format averages \citep[see also][]{klein2024d}. There is no contradiction: that drive is clocked externally - its phase $\beta\tau$ advances smoothly no matter what the inner orbit does. The half-turn (Equation \ref{eq:monodromy}) enters only the kernel: the vector state of a librating cycle repeats after $2T_e$, not $T_e$ (Section \ref{sec:monodromy}), so the resonances of \citepaperIII sit on harmonics of the doubled period - and at resonance the drive's phase advance per cycle compensates the kernel's alternation, so consecutive cycles transfer with the same sign: nothing cancels. The demotion derived here is specific to drives carrying the internal clock $\Omega_e$, whose per-cycle advance is the half-turn itself - the static octupole potential being the prime example.

We close with a remark on phase space. The sign of $C_K$ is set by the angles alone (Equation \ref{eq:CK}), so at the high inclinations relevant for EKL the librating class occupies an $O\left(1\right)$ fraction of phase space: at $i_0=90^\circ$, a uniform distribution of $\omega_0$ places $56$ per cent of systems at $C_K<0$. Whether that share is populated depends on the initial eccentricity. Population syntheses of compact-object triples draw $\omega_0$ uniformly and the inner eccentricity from broad distributions \citep[e.g.][]{antonini2012,stephan2016,hoang2018,shariat2023wd}, and so start a corresponding fraction of their high-inclination systems on librating cycles with finite $\left|C_K\right|$. Planetary syntheses draw $\omega_0$ uniformly as well but typically start nearly circular \citep[e.g.][]{naoz2012,petrovich2015,anderson2016}, which pins $\left|C^0_K\right|\leq\frac{3}{2}e^2_0\ll1$ regardless of the angles: born on the boundary between the classes, these systems surrender the sign of $C_K$ to the octupole within a cycle - and the boundary trajectories followed by \citekdm settled back into rotation within a few cycles - so the near-circular ensembles belong, in effect, to the rotating class. Eccentric starts, however, have been entering the planetary ensembles as well: already \citet{naoz2012} included a single Rayleigh-distributed run, and in the systematic comparison of \citet{weldon2025} the broad draws - the recipe of the compact-object syntheses - reconstruct the observed cold-Jupiter eccentricities better than the circular start. The librating class proper is thus seeded by eccentric inner orbits - and, transiently, by rotating trajectories crossing $C_K=0$ when driven to $\left|j_z\right|\to0$ (\citekdm) - while the analytic theory interpreting all these ensembles descended from the rotating side alone, its examples and parameter surveys initialized at $\omega_0=0$ (\citekdm; e.g. \citealt{lithwick2011,teyssandier2013,lei2022}), which by Equation \ref{eq:CK} forces $C_K=e^2_0>0$. This Letter removes the asymmetry: systems in the librating class now carry amplitudes, timescales and flip windows as analytically calculable as their rotating counterparts.

\section*{Acknowledgements}

We thank Boaz Katz for useful feedback.

\section*{Data Availability}

The codes used in this article will be shared on reasonable request.


\bibliographystyle{mnras}
\bibliography{librating_octupole_is_a_pendulum}


\appendix

\section{The per-cycle coefficients}\label{app:coefficients}

The coefficients of the map (Equation \ref{eq:map}) are computed by linear response around the exact $j_z=0$ cycle of Section \ref{sec:monodromy}. The out-of-plane components ($e_y$ and $j_x$ in the frame of Section \ref{sec:monodromy}) combine into one complex variable whose unforced evolution over one cycle is multiplication by exactly $-1$ - the half-turn of Equation \ref{eq:monodromy} - and variation of parameters reduces each of the first-order coefficients $\chi$, $F$ and $c_\delta$ to a single integral of Jacobi functions over the cycle; the second-order kick $\chi_2$ is the once-iterated response.

Of the four, the kick alone was computed independently in Section \ref{sec:kick}, as the cycle average of a static kernel; the cycle integral reproduces it, in a form that makes the elementary character explicit - the $K\left(k\right)$ in $T_e$ (Equation \ref{eq:Te}) cancels against the $1/K\left(k\right)$ in $\left\langle f_j\right\rangle$ (Equation \ref{eq:fj_librating}):
\begin{equation}
    \chi\left(C_K\right) = \frac{5\sqrt{k}}{8\sqrt{10}}\left[\left(7-14C_K\right)A\left(k\right) - \left(2C_K+3\right)\frac{A\left(k\right)-\sqrt{1-k}}{k}\right].
    \label{eq:kappa_closed}
\end{equation}
with $A\left(k\right)=\arcsin\left(\sqrt{k}\right)/\sqrt{k}$ as after Equation \ref{eq:fj_librating}. The quadrupole slip - the $Fj_z$ residual of the azimuth advance past the geometric $\pi$ in Equation \ref{eq:map} - closes as Equation \ref{eq:F}. The octupole slip $c_\delta$ is an exact one-cycle integral of the linear response, collecting two channels: the direct projection of the octupole's out-of-plane forcing of $\mathbf{e}$, and the feedback of the kick accumulating within the cycle through the quadrupole precession channel. The direct part dominates and closes in elementary form,
\begin{equation}
    c^\text{dir}_\delta=\frac{\sqrt{5}\left(13-10C_K\right)}{8\sqrt{2C_K+3}},
    \label{eq:c_delta_dir}
\end{equation}
and the feedback remainder is a single correlation integral, evaluated numerically. The second-order kick $\chi_2$ has no such integral; it is extracted from a one-time calibration: the full DA equations are integrated across a grid of $C_K$ spanning the window - at a single $\epsilon_\text{oct}=10^{-3}$, $12$ initial azimuths per point, $40$ cycles each - the per-cycle increments of $\left(j_z,\Omega_e\right)$ are read off at the eccentricity extrema, and linear fits to the map (Equation \ref{eq:map}) return its coefficients: $\chi_2$ is the $\sin2\Omega_e$ amplitude of the $j_z$ line ($\chi$ is recovered to $10^{-5}$, the amplitudes forbidden by the half-turn parity fit to zero, and the fitted slips match $c^\text{dir}_\delta$ plus the feedback remainder to four decimal places). Since the coefficients are functions of $C_K$ alone, the calibration is done once, independent of $\epsilon_\text{oct}$ and of any application. At the deep edge $\chi_2$ is pinned in closed form by a degeneracy identity: as $C_K\to-\frac{3}{2}$ the cycle degenerates onto the Kozai fixed point and the slow torque must die with it, $\Lambda\left(-\frac{3}{2}\right)=0$, while the closed forms give $\chi\left(\chi F+c_\delta\right)\to\frac{49}{12}\sqrt{15}$ - Equation \ref{eq:Lambda} then forces $\chi_2\left(-\frac{3}{2}\right)=-\frac{49}{48}\sqrt{15}$, which the measured values approach.

\begin{figure}
 \begin{centering}
 \includegraphics[width=\columnwidth]{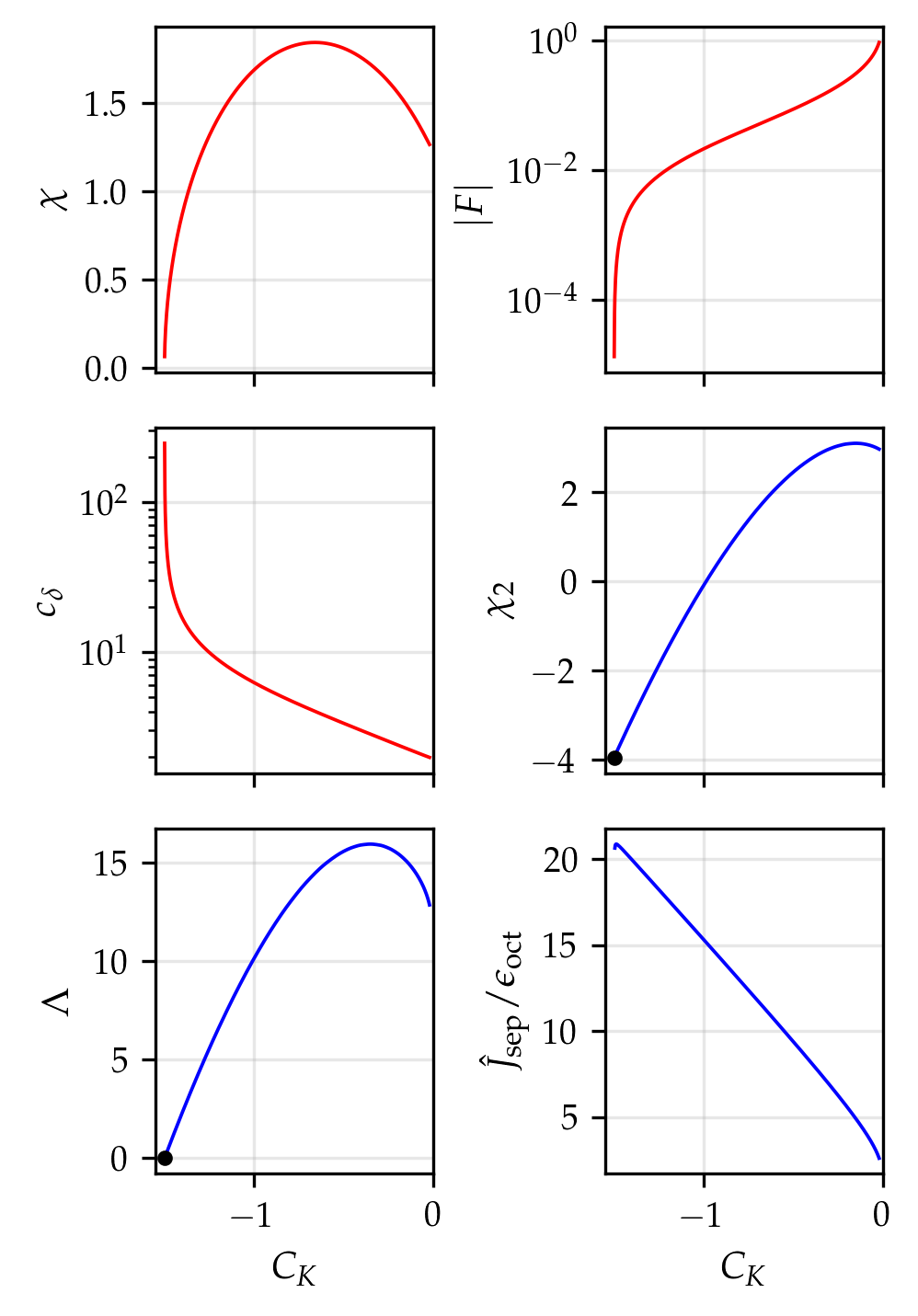}
 \par\end{centering}
 \caption{The coefficients of the per-cycle map and of the slow pendulum across the librating window. In red, the closed forms: the kick $\chi$ (Equations \ref{eq:Te}, \ref{eq:kick}-\ref{eq:fj_librating} and \ref{eq:kappa_closed}), the quadrupole slip $\left|F\right|$ (Equation \ref{eq:F}) and the octupole slip $c_\delta$ (Equation \ref{eq:c_delta_dir} plus the numerically evaluated feedback remainder). In blue, the numerically extracted second-order kick $\chi_2$ and the quantities inheriting it: the torque $\Lambda$ (Equation \ref{eq:Lambda}, vanishing at the edge) and the separatrix amplitude $\Jh_\text{sep}/\epsilon_\text{oct}$ (Equation \ref{eq:Jsep}). Black dots mark the analytic edge values $\chi_2\left(-\frac{3}{2}\right)=-\frac{49}{48}\sqrt{15}$ and $\Lambda\left(-\frac{3}{2}\right)=0$.\label{fig:coefficients}}
\end{figure}

The structural origin of these closed forms is a duality of the simple pendulum (\citealt{basha2025}): the same trajectory admits two half-angle descriptions, exchanged by $k\leftrightarrow\frac{1}{k}$ - one circulating, one swinging. For $C_K<0$ the circulating description is the polar angle of $\mathbf{e}$ in the plane perpendicular to the $\mathbf{j}$-line of Section \ref{sec:monodromy}: it advances by $\pi$ per eccentricity cycle (Equation \ref{eq:monodromy}), and the cycle integrals above are taken with respect to it. The feedback remainder, the one piece left to numerical evaluation, is the correlation of the two descriptions against each other. The coefficients are plotted across the librating window in Figure \ref{fig:coefficients}.

\bsp	
\label{lastpage}
\end{document}